# Infinite-server queueing MAPk|Gk|∞ model with k Markov arrival streams, random volume of customers in random environment subject to catastrophes

K. Kerobyan [1], R. Kerobyan [2], K. Enakoutsa[1]
[1] khanik.kerobyan@csun.edu , [2] rkerobya@ucsd.edu
*[1]California State University Northridge, CSUN, Northridge, USA;*
*[2]University of California San Diego, UCSD, San Diego, USA*

In this paper the infinite server queue model MAPk|Gk|∞ in semi-Markov random environment with k Markov arrival streams, random resources of customers, and catastrophes is considered. After catastrophes occur, all customers in the model are flashed out and the system jumps into recovery station. After the recovery time the model works from the empty state. The transient and stationary joint distributions of numbers of different types of customers in the model at moment t, numbers of different types of served in interval [0,t) customers, volume of accumulated resources in the model at moment t, and total volume of served resources in [0,t) for the model without catastrophes are found. The transient and stationary joint distributions of numbers of different types of customers in the model at moment t, and volume of accumulated resources in the model at moment t and their moments for the model with catastrophes are obtained. All results are obtained using Danzig's collective marks method and renewal theory methods.

**Introduction**

The distinguishing characteristic of New Generation Networks (NGN) is the integration of heterogeneous resources, applications, technologies, customers and data into one united information infrastructure which is ubiquitous and accessible anytime and anywhere. This integration process includes all layer of NGN and makes its QoS metrics guaranteeing more challenging [1]. To solve the NGN optimal design and performance providing problems the methods of statistical simulation and modeling are widely used. However the application of these methods and tools even for several elements of NGN (for example a protocol, a server, or a canal) is complicated because of networks statistical processes (traffic and service) nature [2]. As shown by large number of measurements the traffic of modern IP networks can be characterized by the heterogeneousness, the non-stationarity, the burstiness, the short-range and the long-range dependence. These factors make the modeling and performance evaluation of modern networks more challenging [3]. Network traffics in queueing theory are generally described by based on finite Markov Processes traffic models: Markov Arrival Process (MAP), Batch Markov Arrival Process (BMAP), Marked Arrival Process (MMAP) and their generalizations [4]. The MMAP and MAP arrivals properties and their applications are presented in [5,6] and are not duplicated here. To evaluate the network canals' performance main parameters: capacity, delay and packet loss probability, the infinity server queue models are widely used.

An infinite-server model M|SM|∞ with a Poisson arrival process and with semi-Markovian (SM) service times. The transient and asymptotic results for PGF of queue-length process is obtained by means of CMM [7]. The infinite-server queue $BM_k|G_k|\infty$ with *k* correlated heterogeneous customers in a batch is studied in [8]. In steady state, the joint PGF of queue

length of customers is derived by using CMM and conditional expectations. The generalization of the model for the queue $BMAP_k|G_k|\infty$ with structured batch arrival of *k* types of customers is considered in [9]. In steady state, the differential equations for PGF of queue length and its moments are obtained. The first and second order asymptotes of queue length for the models MAP|G|∞, MMPP|G|∞, G|G|∞ based on supplementary variable method are studied in [10].

To evaluate the impact of network environment on networks performance metrics the infinite-server models in the random enviroment (RE) are applied. The queue size distribution of the model M|G|∞ with semi-Markov (SM) environment under asymptotic condition of high arrival rate and frequent environment transitions is studied in [11]. By the method of supplementary variable and the original method of dynamic screening first and second order asymptotes of queue size distribution are obtained. The queue M|G|∞ in random environment with clearing mechanism is studied in [12]. The environmental clearing process is modeled by an *m*-state irreducible SMP. The transient and steady-state queue length distributions by using renewal arguments are obtained. The $MMAP_k|G_k|\infty$ queue in SM environment and catastrophes is studied in [13]. The PGFs of joint distributions of queue size and number of served customers by using renewal arguments and differential equations are found.

In many applications of queue models such as computer and communication networks, transportation systems customers characterize by the vector of requesting resources which components can be random quantities. Despite the importance of the queueing models and their applications, there are very few works devoted to research of queue resource models, see e.g. [13-19].

The main methods to study the infinite-server queues are: supplementary variables method [20], the method which based on properties of exponential distribution and conditional expectations [21], and Danzig's collective marks method (CMM) method [22, 23]. The last method is also is called "supplementary event or catastrophes" method [22] and has been used successfully for queue models with priorities [23]. CMM have been used in [8] for infinite-server model with Poisson arrival of batches. In [7] the method is mentioned but does not used to obtain some results.

In this paper we consider some generalizations of [12,13,15] results for infinite-server $MAP_k|G_k|\infty$ queue. The transient and stationary joint distributions of numbers of different types of customers in the model at moment t, numbers of different types of served in interval [0,t) customers, volume of accumulated resources in the model at moment t, and total volume of served resources in [0,t) for the model without catastrophes are found. The transient and stationary joint distributions of numbers of different types of customers in the model at moment t, and volume of accumulated resources in the model at moment t and their moments for the model with catastrophes are obtained. All results are obtained using Danzig's collective marks method and renewal theory methods.

1. **Model Description.**

We consider first an infinite-server $MAP_k|G_k|\infty$ model in random environment (RE) with *K* types of customers without catastrophes. The RE operates according to stationary, irreducible semi-Markov process (SMP) $\xi_0(t)$, $t \geq 0$ with finite state space $S = \{1, 2, ..., k\}$. The SMP is given by the vector of initial distribution $\{p_i^0, i \in S\}$ and SM matrix $T(t) = \|T_{ij}(t)\|$, $t \geq 0, i, j \in S$. Customers arrive according to homogeneous MAPs which are given by characteristic matrices $\{D_{0r}, D_{1r}, 1 \leq r \leq K\}$. $D_{0r}$ is a non-singular matrix with negative diagonal elements and $D_{1r}$ is a non-negative matrix. The phase process (PP) $J_r(t)$ of *r*-th MAP is an irreducible Markov process (MP) with generating matrix $D_r$ and finite set of states $E_r$. $D_r$ is a matrix of $m_r \times m_r$ size.

$$D_r = D_{0r} + D_{1r}, \ D_r e_r = 0, \ \pi_r D_r = 0, \ \pi_r e_r = 1, 1 \leq r \leq K,$$

where $e_r$ is a unit column vector, and $\pi_r$ is the vector of stationary distribution $\pi_r = (\pi_1,...,\pi_{m_r})$ of PP $J_r(t)$.

Instead of $K$ independent MAPs we will consider the superposed arrival process. It is known [6] that superposed process is a MAP as well. Let $m$ be $\prod_{r=1}^{K} m_r$. To distinguish arrivals of one type from the others, we introduce the following $m \times m$ matrices;

$$D_0 = D_{01} \oplus D_{02} \oplus ... \oplus D_{0K}, \ D_r = I_1 \otimes ... \otimes I_{r-1} \otimes D_{1r} \otimes I_{r+1} \otimes ... \otimes I_K, \ r = 1,...,K,$$

$$D = D_0 + D_1, \ De = 0, \ \pi D = 0, \ \pi e = 1, \ \pi = \pi_1 \otimes \pi_2 \otimes ... \otimes \pi_K,$$

where $\otimes$ (resp. $\oplus$) denotes the Kronecker product (resp. the Kronecker sum) and $I_r$ denotes the identity matrix of order $m_r$, $\pi$ is a vector of size $m$.

For superposed MAP (SMAP) the stationary arrival rate of customers is given by

$$\lambda = \sum_{r=1}^{K} \pi D_r e = \sum_{r=1}^{K} \lambda_r, \ \lambda_r = \pi D_r e.$$

The service of arriving customers begins immediately. Let the random variable (r.v.) $\tau_r$ be a service time of $r$ type customers, and $\tau = (\tau_1, \tau_2, ..., \tau_K)$ is a vector of service times. Components of $\tau$ are i.i.d. r.v.s which depend on only on the type of the customer and state of environmental SMP. R.v.s $\tau_r$ have general distribution $G_r(t) = P(\tau_r < t)$ and finite mean value $\bar{\tau}_r, 1 \leq r \leq K$.

Each arriving and departing $r$ type customer characterized by $k$-dimensional volume (resource) vector $\zeta_r = (\zeta_{1r},...,\zeta_{kr})$ and $\sigma_r = (\sigma_{1r},...,\sigma_{kr})$ resp., which have non-negative components, $1 \leq r \leq K$. Let $C_r(x) = P(\zeta_{1r} \leq x_1,...,\zeta_{kr} \leq x_k)$ and $G_r(x) = P(\sigma_{1r} \leq x_1,...,\sigma_{kr} \leq x_k)$ be the joint distributions of resource vectors $\zeta_r$ and $\sigma_r$, where $x = (x_1, x_2,...,x_k)$. We assume that the service time vector $\tau$ and the resource vectors $\zeta = (\zeta_1, \zeta_2,...,\zeta_K)$, $\sigma = (\sigma_1, \sigma_2,...,\sigma_K)$ are mutually independent.

When SMP $\xi(t), t \geq 0$ jumps from state $i$ to the state $r$ all customers in the model are instantly flashed out and the model jumps into empty state. Let consider the related SMAP counting processes (CP) $N(t), N_s(t), M(t) : N(t) = (N_1(t),..,N_K(t))$, $N_s(t) = (N_{1s}(t),..,N_{Ks}(t))$, $M(t) = (M_1(t),..,M_K(t))$, where $N_r(t)$ and $M_r(t)$ are the number of $r$ type customers arriving and serving in time interval $[0,t)$, and $N_{rs}(t)$ is a number of $r$ type customers being in service at the time $t$. Let $\beta(t) = (\beta_1(t),...,\beta_K(t))$ and $\alpha(t) = (\alpha_1(t),...,\alpha_K(t))$ be the vectors of total resource served during the time interval $[0,t)$, and accumulated in the model at the time $t$, resp.. The components of $\beta(t)$, $\alpha(t)$, $N_s(t)$ and $M(t)$ vectors are defined as $\beta_r(t) = \sum_{i=1}^{M_r(t)} \sigma_{ri}$, $\alpha_r(t) = \sum_{i=1}^{N_r(t)} \zeta_{ri}$.

Suppose that at the initial time $t = 0$ the model is empty, $N(0) = M(0) = 0$.

## 2. The counting process.

Let consider the CP $\{N(t), J(t); t \geq 0\}$ with matrix $P(n,t)$, $n = (n_1,...,n_k)$, of transition probabilities: $P(n,t) = \| p_{ij}(n,t) \|$, $p_{ij}(n,t) = P(N(t) = n, J(t) = j \mid J(0) = i)$, $1 \leq i, j \leq m$.

Let define the following generating functions (GF) $D(z)$, $P(z,t)$

$$D(z) = D_0 + \sum_{r \geq 1} z_r D_r, \ |z_r| \leq 1, \ 0 \leq r \leq K, \ P(z,t) = \sum_{n \geq 0} z^n P(n,t),$$

where $z = (z_1, z_2,...,z_k)$ and $z^n = (z_1^{n_1}, z_2^{n_2},...,z_k^{n_k})$.

**Theorem 1.** The PGF of counting process $\{N(t), J(t); t \geq 0\}$ $P(z,t)$ satisfies the *basic differential equation*

$$\frac{\partial}{\partial t} P(z,t) = D(z) P(z,t), \quad |z| \leq 1,\tag{1}$$

with initial conditions $P(z,0) = 1$.

The solution of differential equation (1) is given by

$$P(z,t) = \exp\{D(z)t\}. \tag{2}$$

**Proof.** The transition probabilities $\{P(n,t), n \geq 0\}$ of CP $N(t)$ satisfy the following Kolmogorov backward differential equations

$$\frac{d}{dt}P(n,t) = P(n,t)D_0 + \sum_{r=1}^{K} D_r P(n-e_r,t), \quad n \geq 0,$$

with initial condition $P_i(n,0) = 0, n > 0$, $P_i(0,0) = 1, i = 1, 2, ..., K$.

Where $e_r = (0, ..., 0, 1, 0, ..., 0)$ is a vector with 1 in r-th position. Pre-multiplying each equation by corresponding $z^n$ after summation we get differential equations for PDF $P(z,t)$. The solution of this equation in matrix exponential form is given by (1).

### 3. Thinning MAP.

Let consider the following Bernoulli thinning process (TP) of MAP. Each type $r$ customer which arrives at the time $t$ can join the main stream by probability $p_r(t)$ and can be ignored by probability $1 - p_r(t)$.

**Lemma.** The thinned process is a MAP which counting process has rate matrix GF $D_T(z,t)$ and PGF $P_T(z,t)$,

$$D_T(z,x) = D_0 + \sum_{r=1}^{K} D_r[1 - p_r(x) + z_r p_r(x)], \tag{3}$$

$$P_T(z,t) = \exp\{\int_0^t D_T(z,x)dx\}.$$

The first part of lemma is a consequence of more general result for Markov-additive processes of arrivals [24]. The resulting thinned MAP characteristic matrices are

$$D_{T0} = D_0 + \sum_{r=1}^{K} D_r[1 - p_r(x)], \quad D_{Tr} = D_r p_r(x), \quad r = 1, 2, ..., K.$$

The subject of our interest is the joint distribution

$$P(n,m,x,y,t) = P(N_s(t) = n, M(t) = m, \alpha(t) \leq x, \beta(t) \leq y).$$

### 4. Model Analysis.

Let suppose that the SMP $\xi_0(t), t \geq 0$ is in state $i \in S$ and consider the dynamic of the the model in interval $[u,t]$. Each $r$ type customer arriving at the time $u$ will be in service at time $t$ by $1 - B_{ri}(t-u)$ probability and will finish its service before time $t$ by $B_{ri}(t-u)$ probability. So the arrival and service of customers in the model can be considered as a special Bernoulli TP with thinning probability $p_{ri}(t) = B_{ri}(t-u)$.

Let $A_{jk}^i(n,m,x,y,u,t)$ be the joint probability that $n$ customers are in service at time $t$, and $m$ customers are already served in $[0,t)$, total resource in the model at time $t$ is $\alpha(t) \leq x$ and total served resource in $[0,t)$ is $\beta(t) \leq y$, PP $J(u)$ is in the phase $j \in E$ under condition that at initial moment $t = 0$ the model was empty, and PP $J(0)$ was in phase $k \in E$.

Let denote by $\tilde{A}^i(z_1, z_2, s_1, s_2, u, t)$ the matrix which elements are Laplace - Stieltjes (LST) and $z$ transformation of $A_{jk}^i(n,m,x,y,u,t)$ and by $\tilde{C}_{ri}(s_1)$, $\tilde{G}_{ri}(s_2)$ denote LST of $C_{ri}(x)$ and $G_{ri}(y)$. For homogeneous model we have $\tilde{A}^i(z_1, z_2, s_1, s_2, t) = \tilde{A}^i(z_1, z_2, s_1, s_2, u, t)$, e.g. see [25].

$$\tilde{C}_{ri}(s_1) = \int_0^\infty e^{-s_1 x} dC_{ri}(x), \quad \tilde{G}_{ri}(s_2) = \int_0^\infty e^{-s_2 y} dG_{ri}(y).$$

$$\tilde{A}^i(z_1, z_2, s_1, s_2, t) = \sum_{n=0}^{\infty} \sum_{m=0}^{\infty} z_1^n z_2^m \int_0^\infty \int_0^\infty e^{-s_1 x - s_2 y} A^i(n, m, dx, dy, t), |z_1| \leq 1, |z_2| \leq 1.$$

By using CMM we can prove the following result

**Theorem 2.** The PGF of the model $MAP_r|G_r|\infty$ $\tilde{A}^i(z_1, z_2, s_1, s_2, t)$ is given by

$$\tilde{A}^i(z_1,z_2,s_1,s_2,t) = \exp\{\int_0^t [D_0(i)+\tilde{S}_i(z_1,z_2,s_1,s_2,u)]du\}, \quad |z_1|\le 1, |z_2|\le 1. \tag{4}$$

where $\tilde{S}_i(z_1,z_2,s_1,s_2,t) = \sum_{r=1}^{K} D_{ri}[z_{2r}\tilde{G}_{ri}(s_2)B_{ri}(t) + z_{1r}\tilde{C}_{ri}(s_1)(1-B_{ri}(t))]$.

**Proof.** The queueing process in the model can be considered as a CP of some special TP. According to theorem 1 the CP of that TP has matrix exponential form for every state of environmental SMP. Let define the rate of that matrix exponential function namely the rate of no *blue* customers. Suppose that a customer of type $r$ arrives at moment $u$ with rate $D_r$. This customer will be served up to moment $t$ with probability $B_r(t-u)$ or will be in the model at moment $t$ with probability $1-B_r(t-u)$. Let mark each served type $r$ customer *red* by $z_{2r}\tilde{G}_r(s_2)$ or *blue* by $1-z_{2r}\tilde{G}_r(s_2)$ with probabilities. Alike, we mark each serving in the model type $r$ customer *red* by $z_{1r}\tilde{C}_r(s_1)$ or *blue* by $1-z_{1r}\tilde{C}_r(s_1)$ with probabilities. Then $D_r[z_{2r}\tilde{G}_r(s_2)B_r(t-u) + z_{1r}\tilde{C}_r(s_1)(1-B_r(t-u))]$ is the rate of *red* type $r$ customers arriving at moment $u$ and the common rate of *red* (all types) customers arriving at moment $u$ is $\tilde{S}_i(z_1,z_2,s_1,s_2,u)$. Finally, the common rate of *red* (all types) customers arriving in $[0,t)$ is

$$\int_0^t [D_0(i)+\tilde{S}_i(z_1,z_2,s_1,s_2,u)]du.$$

Recall, that $D_0(i)t$ is no customer arrivals' rate in $[0,t)$ interval.

**Theorem 3.** The PGF $\tilde{A}^i(z_1,z_2,s_1,s_2,t)$ satisfy the following basic differential and integral equations

$$\tilde{A}^i(z_1,z_2,s_1,s_2,t) = e^{D_0(i)t} + \int_0^t e^{D_0(i)u}\tilde{S}_i(z_1,z_2,s_1,s_2,u)\tilde{A}^i(z_1,z_2,s_1,s_2,t-u)du. \tag{5}$$

$$\frac{\partial}{\partial t}\tilde{A}^i(z_1,z_2,s_1,s_2,t) = [D_0(i)+\tilde{S}_i(z_1,z_2,s_1,s_2,t)]\tilde{A}^i(z_1,z_2,s_1,s_2,t), \ i\in S, \tag{6}$$

with initial conditions $\tilde{A}^i(z_1,z_2,s_1,s_2,0) = I$,

where $\tilde{S}_i(z_1,z_2,s_1,s_2,t) = \sum_{r=1}^{K} D_{ri}[z_{2r}\tilde{G}_{ri}(s_2)B_{ri}(t) + z_{1r}\tilde{C}_{ri}(s_1)(1-B_{ri}(t))]$.

When $z_2=1, s_2=0$ from (4) - (6) we obtain the PGF of joint distribution number of busy servers and total resources in the model $\bar{\tilde{R}}^i(z_1,s_1,t) = \tilde{A}^i(z_1,1,s_1,0,t)$.

## 5. MAP$_K$|G$_K$|∞ model with catastrophes.

Let consider the general homogeneous Markovian model under influence of SMP generated catastrophes. After every transition of environmental SMP all customers are flashed out of the model, then the model jumps into the failure state. The repair times $\vartheta_i$ of the failed model are i.i.d. r.v. according to general distribution $U_i(t) = P(\vartheta_i \le t)$ with finite mean value $\bar{\vartheta}_{1i}$. After a repair of the model is completed, the model returns to empty state (0-state) and begins to work from that state. Let define a new SMP $\xi(t)$ with SM matrix $Q(t) = \|Q_{ij}(t)\|$ in finite state space $S$. The elements of SM matrix $Q_{ij}(t)$ are define as convolution of $T_{ij}(t)$ and DF $U_j(t)$: $Q_{ij}(t) = T_{ij}(t) * U_j(t)$ $i,j\in S$. When the SMP is in $i$-th state all parameters of the model are related to that state: DF of inter-arrival time of customers, DF and rates of service time of customers, their resource vectors. Let $\bar{\tilde{P}}(n,m,s_1,s_2,t,i)$ and $\tilde{P}(n,m,s_1,s_2,t,i)$ defined the LST of probabilities of having in the model $n=(n_1,n_2,...,n_k)$ customers at the moment $t$ and having $m=(m_1,m_2,...,m_k)$ served customers in $[0,t)$ when environmental SMP is in state $i$, for the models without catastrophes and with catastrophes, respectively. The following theorem gives the connection between these two probabilities,

**Theorem 4.**

$$\tilde{P}(n,m,s_1,s_2,t,i) = (1-F_i(t))\tilde{\tilde{P}}(n,m,s_1,s_2,t,i) + \sum_{j\in S}\int_0^t \tilde{P}(n,m,s_1,s_2,t-u,j)dQ_{ij}(u). \quad i\in S. \qquad (7)$$

The solution of the equations can be found as

$$\tilde{P}(n,m,s_1,s_2,t,i) = \bar{F}_i(t)\tilde{\tilde{P}}(n,m,s_1,s_2,t,i) + \sum_{j\in S}\int_0^t \bar{F}_j(t-u)\tilde{\tilde{P}}(n,m,s_1,s_2,t-u,j)dH_{ij}(u). \quad i\in S, \qquad (8)$$

where $\bar{A}(t) = 1 - A(t)$, $F(t) = \{F_i(t), i\in S\}$ is a sojourn time distribution vector of SMP

$$F_i(t) = \sum_{j\in S} T_{ij}(t), \; i\in S,$$

$H(t) = \|H_{ij}(t)\|$ is a renewal matrix of SMP which components satisfy the following equations

$$H_{ij}(t) = 1 - \sum_{k\in S} Q_{ik}(t) + \sum_{k\in S}\int_0^t H_{kj}(t-u)dQ_{ik}(u). \quad i,j\in S. \qquad (9)$$

The proof of (7), (8) can be done by using standard renewal arguments (see for example [22-26]).

Let $\tilde{P}(z_1,z_2,s_1,s_2,t,i)$ and $\tilde{\tilde{P}}(z_1,z_2,s_1,s_2,t,i)$ are the PGFs of $\tilde{P}(n,m,s_1,s_2,t,i)$ and $\tilde{\tilde{P}}(n,m,s_1,s_2,t,i)$ resp.. Then we get from (7), (8)

$$\tilde{P}(z_1,z_2,s_1,s_2,t,i) = (1-F_i(t))\tilde{\tilde{P}}(z_1,z_2,s_1,s_2,t,i) + \sum_{j\in S}\int_0^t \tilde{P}(z_1,z_2,s_1,s_2,t-u,j)dQ_{ij}(u), \quad i\in S, \qquad (10)$$

which solution is

$$\tilde{P}(z_1,z_2,s_1,s_2,t,i) = \bar{F}_i(t)\tilde{\tilde{P}}(z_1,z_2,s_1,s_2,t,i) + \sum_{j\in S}\int_0^t \bar{F}_j(t-u)\tilde{\tilde{P}}(z_1,z_2,s_1,s_2,t-u,j)dH_{ij}(u), \quad i\in S, \qquad (11)$$

The proof can be done by using CMM or standard renewal argument (see for example [23]). Let consider the proof by CMM. We mark the customers as in theorem 2 case in *red* or *blue* color resp. Then left side of (10) $\tilde{P}(z_1,z_2,s_1,s_2,t,i)$ is a probability of event "in the model with catastrophes there are no *blue* customers at moment $t$ and the SMP is in state $i$". This event can be realized either "the SMP is in state $i$ at moment $t$ given that at initial time $t=0$ it was in state $i$ and in the model without catastrophes there are no *blue* customers in interval $[0,t)$" or "SMP jumps from initial state $i$ into state $j$ in $[u,du)$", $u\leq t$ (with probability $dQ_{ij}(u)$), the model jumps into 0-state and from that state "in the model are not *blue* customers in interval $[t-u,t)$" (with probability $\tilde{P}(z_1,z_2,s_1,s_2,t-u,j)$). In the proof of (11) we have to paraphrase second term of right side: "SMP jumps from $i$ state into state $j$ in $[u,du)$", $u\leq t$ (with probability $dH_{ij}(u)$, the model jumps into 0-state and "SMP will stay in the state $j$ during interval of time $[t-u,t)$ and in the model without catastrophes starting from 0-state (empty state) are not *blue* customers in interval of time $[t-u,t)$" (with probability $\bar{F}_j(t-u)\tilde{\tilde{P}}(z_1,z_2,s_1,s_2,t-u,j)$).

**Theorem 5.** The limiting distributions of $\tilde{P}(n,m,s_1,s_2,t,i)$ and $\tilde{P}(z_1,z_2,s_1,s_2,t,i)$ have the form

$$\tilde{P}(n,m,s_1,s_2) = \lim_{t\to\infty}\sum_{i\in S} p_i^0 \tilde{P}(n,m,s_1,s_2,t,i) = \sum_{j\in S}\frac{q_j}{\bar{\eta}_j}\int_0^\infty (1-F_j(u))\tilde{\tilde{P}}(n,m,s_1,s_2,u,j)du, \; i\in S, \qquad (12)$$

$$\tilde{P}(z_1,z_2,s_1,s_2) = \lim_{t\to\infty}\sum_{i\in S} p_i^0 \tilde{P}(z_1,z_2,s_1,s_2,t,i) = \sum_{j\in S}\frac{q_j}{\bar{\eta}_j}\int_0^\infty (1-F_j(u))\tilde{\tilde{P}}(z_1,z_2,s_1,s_2,u,j)du, \; i\in S, \qquad (13)$$

where $\bar{\eta}_i = \int_0^\infty (1-\sum_{j\in S} Q_{ji}(u))du$, $q_i = \dfrac{\bar{\eta}_i \rho_i}{\sum_{r\in S}\bar{\eta}_r \rho_r}$, $\sum_{r\in S} q_r = 1$, $\rho_i = \sum_{r\in S} p_{ri}\rho_r$, $\sum_{r\in S}\rho_r = 1$, $p_{ri} = Q_{ri}(\infty)$, $r,i\in S$.

Let $\tilde{f}(s)$ denote the Laplace Transformation of a function $f(x)$, $\tilde{f}(s) = \int_0^\infty e^{-su}f(u)du$.

**Corollary.** When $F_i(t) = 1 - e^{-\nu_i}$, $i\in S$ then for LT of $\tilde{P}(n,m,s_1,s_2,s,i)$, PGF $\tilde{P}(z_1,z_2,s_1,s_2,s,i)$ and their limiting values we get

$$\tilde{P}(n,m,s_1,s_2,s,i) = \tilde{\tilde{P}}(n,m,s_1,s_2,s+\nu_i,i) + \frac{1}{s}\sum_{j\in S}\tilde{\tilde{P}}(n,m,s_1,s_2,s+\nu_j,j)\tilde{H}_{ij}(s). \quad i\in S \qquad (14)$$

$$\tilde{P}(z_1,z_2,s_1,s_2,s,i) = \tilde{\tilde{P}}(z_1,z_2,s_1,s_2,s+v_i,i) + \frac{1}{s}\sum_{j \in S} \tilde{\tilde{P}}(z_1,z_2,s_1,s_2,s+v_j,j)\tilde{H}_{ij}(s). \quad i \in S \tag{15}$$

$$\tilde{P}(z_1,z_2,s_1,s_2) = \sum_{i \in S} q_i v_i \tilde{\tilde{P}}(z_1,z_2,s_1,s_2,v_i,i), \quad \tilde{P}(n,m,s_1,s_2) = \sum_{i \in S} q_i v_i \tilde{\tilde{P}}(n,m,s_1,s_2,v_i,i).$$

Let consider the infinite-server models $MAP_k|G_k|\infty$ with catastrophes. Assume that $\tilde{P}(z_1,s_1,t)$ is a LST of PGF joint distribution of number of busy servers and total accumulated resources in the model at moment $t$ given that at $t=0$ SMP is in state $i$ for the model with catastrophes and $\tilde{P}(z_1,s_1)$ is its limit value.

**Theorem 6.** The PGFs $\tilde{P}(z_1,s_1,t)$ and its limit value $\tilde{P}(z_1,s_1)$ are given by

$$\tilde{P}(z_1,s_1,t) = \sum_{i \in S} p_i^0 \tilde{P}(z_1,s_1,t,i) \tag{16}$$

where $\{p_i^0, i \in S\}$ is a vector of initial distribution of SMP, and $\tilde{P}(z_1,s_1,t,i)$ satisfies the following integral equations

$$\tilde{P}(z_1,s_1,t,i) = e^{\int_0^t [D_0(i)+\tilde{S}_i(z_1,s_1,u)]du}(1-F_i(t)) + \sum_{r \in S} \int_0^t \tilde{P}(z_1,s_1,t-u,r)dQ_{ir}(u), \quad i \in S, \tag{17}$$

which solutions are

$$\tilde{P}(z_1,s_1,t,i) = e^{\int_0^t [D_0(i)+\tilde{S}_i(z_1,s_1,u)]du}\bar{F}_i(t) + \sum_{r \in S} \int_0^t \bar{F}_r(t-u)e^{\int_0^{t-u}[D_0(r)+\tilde{S}_r(z_1,s_1,x)]dx}dH_{ir}(u), \quad i \in S, \tag{18}$$

$$\tilde{P}(z_1,s_1) = \lim_{t\to\infty}\sum_{i\in S} p_i^0 \tilde{P}(z_1,s_1,t,i) = \sum_{i \in S} \frac{q_i}{\eta_i}\int_0^\infty (1-F_i(x))e^{\int_0^x [D_0(i)+\tilde{S}_i(z_1,s_1,u)]du}dx, \quad i \in S. \tag{19}$$

**Corollary.** From above model when catastrophes occur according to Poisson distribution with parameter $v$ then by (18) we get the results for homogeneous model [25]. For example, for $\tilde{P}(z_1,s_1)$ and $\tilde{P}(z_1,s_1,s)$ we obtain

$$\tilde{P}(z_1,s_1,s) = \tilde{\tilde{P}}(z_1,s_1,s+v)[1+\frac{v}{s}], \quad \tilde{P}(z_1,s_1) = v\tilde{\tilde{P}}(z_1,s_1,v). \tag{20}$$

Where $\tilde{\tilde{P}}(z_1,s_1,v) = \int_0^\infty e^{-vx} e^{\int_0^x [D_0(i)+\tilde{S}_i(z_1,s_1,u)]du}dx, \quad \tilde{\tilde{P}}(z_1,s_1,s+v) = \int_0^\infty e^{-(v+s)x} e^{\int_0^x [D_0(i)+\tilde{S}_i(z_1,s_1,u)]du}dx.$

This result can be interpreted by the CMM. Let consider flow of event $A$ which has exponentially distributed inter-event time with parameter $s$.

First, let (20) write in the form $s\tilde{P}(z_1,s_1,s) = \tilde{\tilde{P}}(z_1,s_1,s+v)[s+v]$, then it can be interpreted as follow: $s\tilde{P}(z_1,s_1,s)$ is a probability of the event "event $A$ appears when in the model with catastrophes no *blue* customers at moment $t$" but it is the same as happens the event "sum of the event $A$ and catastrophes appears when in the model without catastrophes no *blue* customers at moment $t$". The probability of this event is $\tilde{\tilde{P}}(z_1,s_1,s+v)[s+v]$.

**Performance measures of the model.**

Let $\omega_{1r}(t)$, $\omega_{1r}$ and $Var_{1r}(t)$, $Var_{1r}$ denote the transient and steady state mean and variance of queue length of type $r$ customers. Then for $\omega_{1r}(t)$ and $\omega_{1r}$ we get

$$\omega_{1r}(t) = \lim_{s_1 \to 0} \tilde{\omega}_{1r}(s_1,t), \quad \tilde{\omega}_{1r}(s_1,t) = \frac{\partial \tilde{P}(z_1,s_1,t)}{\partial z_{1r}}\bigg|_{z_{1r}=1,z_{11}=z_{12}=...=z_{1k}=1},$$

$$\omega_{1r} = \lim_{s_1 \to 0} \tilde{\omega}_{1r}(s_1), \quad \tilde{\omega}_{1r}(s_1) = \frac{\partial \tilde{P}(z_1,s_1)}{\partial z_{1r}}\bigg|_{z_{1r}=1,z_{11}=z_{12}=...=z_{1k}=1},$$

$$\tilde{\omega}_{1r}(s_1,t) = \sum_{i \in S} p_i^0 \bar{F}_i(t)\tilde{\omega}_{1r}(s_1,t,i) + \sum_{j \in S} \int_0^t \bar{F}_j(t-u)\tilde{\omega}_{1r}(s_1,t-u,j)dH_j(u),$$

$$\tilde{\omega}_{1r}(s_1) = \sum_{j \in S} \frac{q_j}{\bar{\eta}_j} \int_0^\infty (1 - F_j(u))\tilde{\bar{\omega}}_{1r}(s_1, u, j) du. \tag{21}$$

Where $\tilde{\bar{\omega}}_{1r}(s_1, t, i)$ is a transient mean queue length of $r$ type customers of the model without catastrophes when the SMP is in state $i$.

The transient and steady state variance of queue length for type $r$ customers $Var_{1r}(t)$ and $Var_{1r}$ we get

$$Var_{1r}(t) = \omega_{2r}(t) + \omega_{1r}(t)(1 - \omega_{1r}(t)), \quad Var_{1r} = \omega_{2r} + \omega_{1r}(1 - \omega_{1r}),$$

$$\omega_{2r}(t) = \lim_{s_1 \to 0} \tilde{\omega}_{2r}(s_1, t), \quad \tilde{\omega}_{2r}(s_1, t) = \left. \frac{\partial^2 \tilde{P}(z_1, s_1, t)}{\partial z_{1r}^2} \right|_{z_{1r}=1, z_{11}=z_{12}=\ldots=z_{1k}=1},$$

$$\omega_{2r} = \lim_{s_1 \to 0} \tilde{\omega}_{2r}(s_1), \quad \tilde{\omega}_{2r}(s_1) = \left. \frac{\partial^2 \tilde{P}(z_1, s_1)}{\partial z_{1r}^2} \right|_{z_{1r}=1, z_{11}=z_{12}=\ldots=z_{1k}=1},$$

$$\tilde{\bar{\omega}}_{2r}(s_1, t) = \sum_{i \in S} p_i^0 (1 - F_i(t)) \tilde{\bar{\omega}}_{2r}(s_1, t, i) + \sum_{j \in S} \int_0^t (1 - F_j(t-u)) \tilde{\bar{\omega}}_{2r}(s_1, t-u, j) dH_j(u),$$

$$\tilde{\omega}_{2r}(s_1) = \sum_{j \in S} \frac{q_j}{\bar{\eta}_j} \int_0^\infty (1 - F_j(u)) \tilde{\bar{\omega}}_{2r}(s_1, u, j) du. \tag{22}$$

Let $\delta_r(t)$, $\delta_r$, $r = 1, 2, \ldots, K$, be the transient and steady-state mean values of accumulated type $r$ resources in the model and $\delta$ be a total accumulated resources in the model.

$$\delta_r(t) = \pi \bar{\delta}_r(t) e, \quad \bar{\delta}_r(t) = \lim_{s_1 \to 0} \left. \frac{\partial \tilde{P}(z_1, s_1, t)}{\partial s_{1r}} \right|_{z_{11}=z_{12}=\ldots=z_{1k}=1},$$

$$\bar{\delta}_r(t) = \sum_{i \in S} p_i^0 (1 - F_i(t)) \bar{\delta}_r(t, i) + \sum_{j \in S} \int_0^t (1 - F_j(t-u)) \bar{\delta}_r(t-u, j) dH_j(u),$$

$$\delta_r = \lim_{t \to \infty} \delta_r(t), \quad \delta_r = \sum_{j \in S} \frac{q_j}{\bar{\eta}_j} \lambda_{jr} \bar{c}_{1r}(j) \int_0^\infty \int_0^u (1 - F_j(u))(1 - B_{jr}(x)) dx du, \tag{23}$$

where $\bar{\delta}_r(t, j) = \lambda_{jr} \bar{c}_{1r}(j) \int_0^t (1 - B_{jr}(x)) dx$, $\bar{c}_{1r}(j)$ is a mean value of DF $C_{jr}(t)$.

$$\delta = \sum_{r=1}^K \delta_r = \sum_{r=1}^K \sum_{j \in S} \frac{q_j}{\bar{\eta}_j} \lambda_{jr} \bar{c}_{1r}(j) \int_0^\infty \int_0^u (1 - F_j(u))(1 - B_{jr}(x)) dx du.$$

If $L_{losr}$ denote the steady state mean number of destroyed type $r$ customers, then

$$L_{losr} = \lim_{s_1 \to 0} \pi \tilde{L}_{losr}(s_1) e, \tag{24}$$

where $\tilde{L}_{losr}(s_1) = \sum_{j \in S} \frac{q_j}{\bar{\eta}_j} \int_0^\infty \tilde{\bar{\omega}}_{1r}(s_1, u, j) dF_j(u)$.

$$L_{losr} = \sum_{j \in S} \frac{q_j}{\bar{\eta}_j} \lambda_{jr} \int_0^\infty \int_0^u (1 - B_{jr}(x)) dx dF_j(u). \tag{25}$$

If $L_{los}$ is the steady-state total mean number of destroyed customers of all types, then

$$L_{los} = \sum_{r=1}^K L_{losr} = \sum_{r=1}^K \sum_{j \in S} \frac{q_j}{\bar{\eta}_j} \lambda_{jr} \int_0^\infty \int_0^u (1 - B_{jr}(x)) dx dF_j(u). \tag{26}$$

Let $L_{qr}$ and $L_q$ be the steady state mean number of type $r$ and all types customers in the model. Then

$$L_{qr} = \pi \tilde{\omega}_{1r} e = \sum_{j \in S} \frac{q_j}{\bar{\eta}_j} \int_0^\infty (1 - F_j(u)) \pi \tilde{\bar{\omega}}_{1r}(u, j) e du = \sum_{j \in S} \frac{q_j}{\bar{\eta}_j} \lambda_{jr} \int_0^\infty \int_0^u (1 - F_j(u))(1 - B_{jr}(x)) dx du,$$

$$L_q = \sum_{r=1}^{K} L_{qr} = \sum_{r=1}^{K} \sum_{j \in S} \frac{q_j}{\overline{\eta}_j} \lambda_{jr} \int_0^\infty \int_0^u (1-F_j(u))(1-B_{jr}(x))dxdu. \qquad (27)$$

Suppose that MAP is defined by following matrices $D_0(i) = -\alpha_i I$, $D_r(i) = \alpha_{ir} I$, $r = 1, 2, .., K$, $i \in S$, where $I$ is an identity matrix. Then for $\overline{\tilde{P}}(n,t,i), P(n,t), P(n), \omega_{1r}(t,i)$ and $L_{losr}$ we obtain

$$\overline{\tilde{P}}(z,s,t,i) = e^{-\int_0^t \sum_{r=1}^k \lambda_{ir}(1-B_{ir}(x))(1-z_{ir}\tilde{C}_{ir}(s))dx}, \quad \overline{P}(n,t,i) = \prod_{r=1}^{K} \frac{a_{ir}(t)^{n_r}}{n_r!} e^{-a_i(t)}, \quad a_i(t) = \sum_{r=1}^{K} a_{ir}(t)$$

$$P(n,t) = \sum_{j \in S} p_i^0 (1-F_i(t)) \prod_{r=1}^{K} \frac{a_{ir}(t)^{n_r}}{n_r!} e^{-a_i(t)} + \sum_{j \in S} \int_0^t (1-F_j(t-u)) \prod_{r=1}^{K} \frac{a_{ir}(t-u)^{n_r}}{n_r!} e^{-a_i(t-u)} dH_j(u),$$

$$P(n) = \sum_{j \in S} \frac{q_j}{\overline{\eta}_j} \int_0^\infty (1-F_j(u)) \prod_{r=1}^{K} \frac{a_{jr}(u)^{n_r}}{n_r!} e^{-a_j(u)} du.$$

$$\omega_1(t) = \sum_{j \in S} p_j^0 (1-F_j(t)) a_j(t) + \sum_{j \in S} \int_0^t (1-F_j(t-u)) a_j(t-u) dH_j(u),$$

$$\omega_1 = \sum_{j \in S} \frac{q_j}{\overline{\eta}_j} \int_0^\infty (1-F_j(u)) a_j(u) du.$$

$$L_{los} = \sum_{j \in S} \frac{q_j}{\overline{\eta}_j} \int_0^\infty a_j(u) dF_j(u).$$

where $a_{ir}(t) = \alpha_{ir} \int_0^t (1-B_{ir}(u))du$, $i \in S$.

### Conclusion

In this paper we consider the infinite-server $MAP_k|G_k|\infty$ queue in random environment with resource vectors of customers, subject to catastrophes. After the catastrophes occur, all customers in the model are flashed out and the system jumps into recovery (repair) station. After the repair the model works from the empty state. The transient and stationary joint distributions of numbers of different types of customers in the model at moment t, numbers of different types of served in interval [0,t) customers, volume of accumulated resources in the model at moment t, and total volume of served resources in [0,t) for the model without catastrophes are found. The transient and stationary joint distributions of numbers of different types of customers in the model at moment t, and volume of accumulated resources in the model at moment t and their moments for the model with catastrophes are obtained. All results are obtained using Danzig's collective marks method and renewal theory methods.

The obtained results may be applied for evaluating the performance metrics, as well as for finding the optimal strategies of managing resources for a wide class of computer systems and networks, whereas the queue $MAP_r|G_r|\infty$ may be used as a model.